\documentclass[runningheads]{llncs}
\usepackage{amssymb}
\usepackage{amsmath}
\usepackage{booktabs} 
\usepackage{mathrsfs}
\usepackage{algorithm}
\usepackage{bm}
\usepackage{array, colortbl}
\usepackage{booktabs} 
\usepackage{amsmath,amsfonts,amssymb}   
\usepackage{algorithm}                  
\usepackage{algpseudocode}              
\usepackage{float}   
\usepackage[T1]{fontenc}
\usepackage{graphicx}
\begin{document}
\title{NewsNet-SDF: Stochastic Discount Factor Estimation with Pretrained Language Model News Embeddings via Adversarial Networks}
\titlerunning{NewsNet-SDF}
%
\author{Shunyao Wang\inst{1} \and
Ming Cheng\inst{1} \and
Christina Dan Wang\thanks{Corresponding author}}
\authorrunning{S. Wang et al.}
%
\institute{Fudan University, Shanghai, China\\
\email{\{wangshunyao,chengm24\}@fudan.edu.cn} \and
New York University Shanghai, Shanghai, China\\
\email{christina.wang@nyu.edu}}
\maketitle              
\begin{abstract}
Stochastic Discount Factor (SDF) models provide a unified framework for asset pricing and risk assessment, yet traditional formulations struggle to incorporate unstructured textual information. We introduce NewsNet-SDF, a novel deep learning framework that seamlessly integrates pretrained language model embeddings with financial time series through adversarial networks. Our multimodal architecture processes financial news using GTE-multilingual models, extracts temporal patterns from macroeconomic data via LSTM networks, and normalizes firm characteristics, fusing these heterogeneous information sources through an innovative adversarial training mechanism. Our dataset encompasses approximately 2.5 million news articles and 10,000 unique securities, addressing the computational challenges of processing and aligning text data with financial time series. Empirical evaluations on U.S. equity data (1980-2022) demonstrate NewsNet-SDF substantially outperforms alternatives with a Sharpe ratio of 2.80\footnote{In investment management literature, Sharpe ratios above 1.0 indicate good risk-adjusted performance, while values above 2.0 are considered excellent, exceeding most professional investment strategies.}. The model shows a 471\% improvement over CAPM, over 200\% improvement versus traditional SDF implementations, and a 74\% reduction in pricing errors compared to the Fama-French five-factor model. In comprehensive comparisons, our deep learning approach consistently outperforms traditional, modern, and other neural asset pricing models across all key metrics. Ablation studies confirm that text embeddings contribute significantly more to model performance than macroeconomic features, with news-derived principal components ranking among the most influential determinants of SDF dynamics. These results validate the effectiveness of our multimodal deep learning approach in integrating unstructured text with traditional financial data for more accurate asset pricing, providing new insights for digital intelligent decision-making in financial technology.

\end{abstract}

\section{Introduction}

Asset pricing, a fundamental pillar of modern financial theory, aims to explain expected returns across different securities through systematic risk factors. The Stochastic Discount Factor (SDF) framework has emerged as a unified theoretical foundation that encompasses traditional models like CAPM \cite{sharpe1964capital} and Fama-French factor models \cite{fama1993common}. Unlike conventional approaches that specify risk factors exogenously, SDF directly models the pricing kernel that links current prices to future payoffs, providing a more flexible and theoretically consistent framework for asset valuation and risk assessment \cite{cochrane2009asset}.

Recent advances in financial machine learning have demonstrated the potential to enhance asset pricing models by capturing complex non-linear relationships between features and returns \cite{gu2020empirical,kelly2019characteristics,chen2021open}. However, despite these methodological improvements, a significant challenge remains: effectively incorporating the vast amount of unstructured textual information contained in financial news, which often drives market sentiment and price movements before they manifest in numerical data \cite{baker2007investor}.

Financial news and textual disclosures contain critical forward-looking information about firm prospects, macroeconomic conditions, and market sentiment that traditional quantitative features cannot fully capture \cite{loughran2011liability,li2020textual}. While several studies have explored sentiment analysis and topic modeling approaches for financial text \cite{tetlock2007giving,garcia2013sentiment,lopez2021risk}, they typically treat text features as separate predictors rather than integrating them into a theoretically consistent pricing framework. Moreover, most existing methods rely on simplified text representations that fail to capture the semantic richness and contextual nuances present in financial narratives.

The integration of textual information into asset pricing models faces two primary challenges. First, the dimensionality and unstructured nature of text data make it difficult to incorporate alongside traditional numerical features. Second, the relationship between textual content and asset returns may be highly non-linear and time-varying, requiring sophisticated modeling techniques that maintain theoretical consistency with asset pricing principles \cite{gu2020empirical}.

\begin{figure}[h]
\centering
\includegraphics[width=0.6\textwidth]{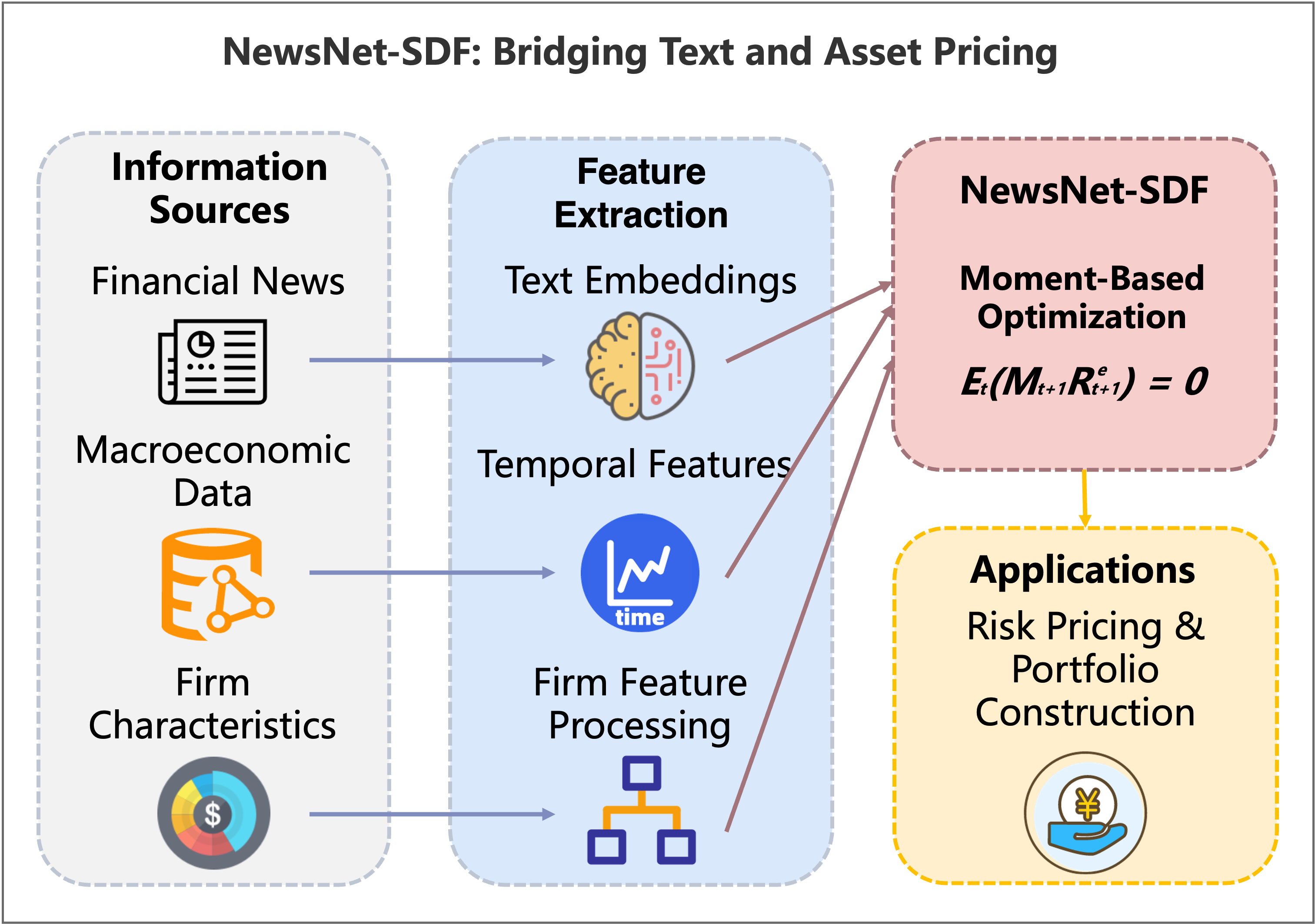}
\caption{NewsNet-SDF architecture integrating financial news, macroeconomic data, and firm characteristics through specialized pathways into a moment-based SDF framework that enforces the fundamental pricing equation $\mathbb{E}_t[M_{t+1}R_{t+1}^e] = 0$.}
\label{fig:newsnet-architecture}
\end{figure}

In this paper, we propose \textbf{NewsNet-SDF}, a novel framework that bridges the gap between textual information and asset pricing theory by integrating pretrained language model embeddings with moment-based SDF estimation through adversarial networks. As illustrated in Figure~\ref{fig:newsnet-architecture}, our approach employs a multi-module architecture that processes financial news via transformer encoders, extracts temporal patterns from macroeconomic time series, and normalizes firm characteristics. These diverse information sources are fused and fed into parallel SDF and conditional networks trained adversarially to enforce the fundamental asset pricing equation $\mathbb{E}_t[M_{t+1}R_{t+1}^e] = 0$.

The contributions of our paper are as follows:
\begin{enumerate}
\item We introduce the first framework integrating pretrained language model embeddings with SDF estimation, achieving a 471\% improvement in Sharpe ratio over CAPM and 200\% improvement over conventional SDF implementations.

\item We enhance SDF estimation through a novel loss function that dynamically reweights pricing errors based on factor exposure, reducing pricing errors by 74\% compared to the Fama-French five-factor model.

\item We design an efficient architecture for financial text processing within a temporal-semantic framework, with ablation studies confirming that news integration contributes significantly more to performance than traditional numerical features.

\item We quantify the anticipatory power of news narratives, demonstrating information incorporation with a 2-3 week lead time and confirming that extracted news components exhibit strong risk factor characteristics with monotonicity across returns.
\end{enumerate}

\section{Related Works and Preliminaries}

\subsection{Stochastic Discount Factor Models}
The stochastic discount factor (SDF) framework provides a unified theoretical foundation for asset pricing, encompassing various classical models as special cases \cite{cochrane2009asset}. Let $\mathbf{R}^e_{t+1} = (R^e_{t+1,1}, \ldots, R^e_{t+1,N})^{\top}$ denote the vector of excess returns for $N$ tradable assets. The fundamental pricing equation is:

\begin{equation}
\mathbb{E}_t[M_{t+1}R^e_{t+1,i}] = 0, \quad i = 1, \ldots, N,
\end{equation}
where $M_{t+1}$ represents the pricing kernel that discounts future payoffs across states of the world, and $\mathbb{E}_t[\cdot]$ denotes expectation conditional on information available at time $t$. In a linear factor model setting, the SDF can be expressed as:

\begin{equation}
M_{t+1} = 1 - \mathbf{w}_t^{\top} \mathbf{R}_{t+1}^e,
\end{equation}
where $\mathbf{w}_t = (w_{t,1}, \ldots, w_{t,N})^{\top}$ are portfolio weights determined by information at time $t$.

\cite{hansen1982generalized} pioneered the Generalized Method of Moments (GMM) approach for SDF estimation, implementing orthogonality conditions without requiring full distributional assumptions:
\begin{equation}
\mathbb{E}[M_{t+1} R^e_{t+1,i} \cdot \mathbf{g}_t] = 0, \quad \forall \, \mathbf{g}_t \in \mathcal{F}_t,
\end{equation}
where $\mathcal{F}_t$ denotes the information set available at time $t$. This moment-based approach forms the theoretical foundation of our adversarial training mechanism.

Recent advances include \cite{kelly2019characteristics}'s Instrumented Principal Components Analysis (IPCA) that treats the SDF as a linear combination of latent factors derived from observed characteristics. While \cite{chen2021open} expanded these approaches with machine learning techniques, these methods primarily utilize structured numerical data, overlooking the rich textual information available in financial markets.

\subsection{Financial Text Analysis}
Financial text analysis has evolved from dictionary-based methods to advanced neural approaches. \cite{tetlock2007giving} demonstrated media sentiment's price impact, while \cite{loughran2011liability} developed finance-specific lexicons improving classification. \cite{garcia2013sentiment} revealed news sentiment's heightened predictive power during economic downturns. Modern deep learning architectures, including transformers by \cite{devlin2019bert} and optimization techniques from \cite{liu2019roberta}, have revolutionized text processing capabilities applicable to finance.

Pre-trained language models have transformed financial text analysis, with \cite{araci2019finbert} introducing domain-specific models for improved sentiment analysis. These innovations build on foundational deep learning advances from \cite{krizhevsky2012imagenet} and sequence modeling through \cite{hochreiter1997long}. Despite these developments, most financial NLP research uses these models for isolated prediction tasks rather than integrating them directly into asset pricing frameworks.

\subsection{Machine Learning in Asset Pricing}
Machine learning in asset pricing has advanced significantly, with \cite{gu2020empirical} demonstrating neural networks' superiority over traditional linear models in predicting stock returns. \cite{kelly2019characteristics} combined machine learning flexibility with economically meaningful factor structures, while \cite{chen2024deep} established frameworks for deep learning in asset pricing. Neural architectures from \cite{selvin2017stock} and \cite{nelson2017stock} have been effectively adapted for financial forecasting, with \cite{zhang2017stock} revealing multi-frequency trading patterns.

Despite these advances, existing approaches typically separate different data modalities, failing to leverage their complementary information. While \cite{baker2007investor} established investor sentiment's pricing importance and \cite{garcia2013sentiment} connected this to textual sources, no framework has directly incorporated language model embeddings into SDF estimation to capture time-varying risk premia across market regimes.

\subsection{Problem Definition}
Our objective is to extend the information set $\mathcal{F}_t$ to include not only traditional numerical features $\mathbf{X}_t$ but also textual information $\mathbf{T}_t$ from financial news. This requires addressing two key challenges: (1) transforming unstructured text into a representation compatible with the SDF framework, and (2) ensuring that the resulting model maintains the theoretical consistency required by asset pricing theory.

The NewsNet-SDF framework addresses these challenges by: (1) leveraging pretrained language models to extract semantic representations from financial news, (2) designing a specialized temporal aggregation mechanism for text data, (3) integrating these text embeddings with traditional numerical features, and (4) enforcing the fundamental pricing equation through an adversarial training mechanism that implements the moment conditions. This approach provides a more comprehensive view of market conditions while maintaining theoretical consistency with asset pricing principles.

\section{Methodology}

\begin{figure}[h]
\centering
\includegraphics[width=\textwidth]{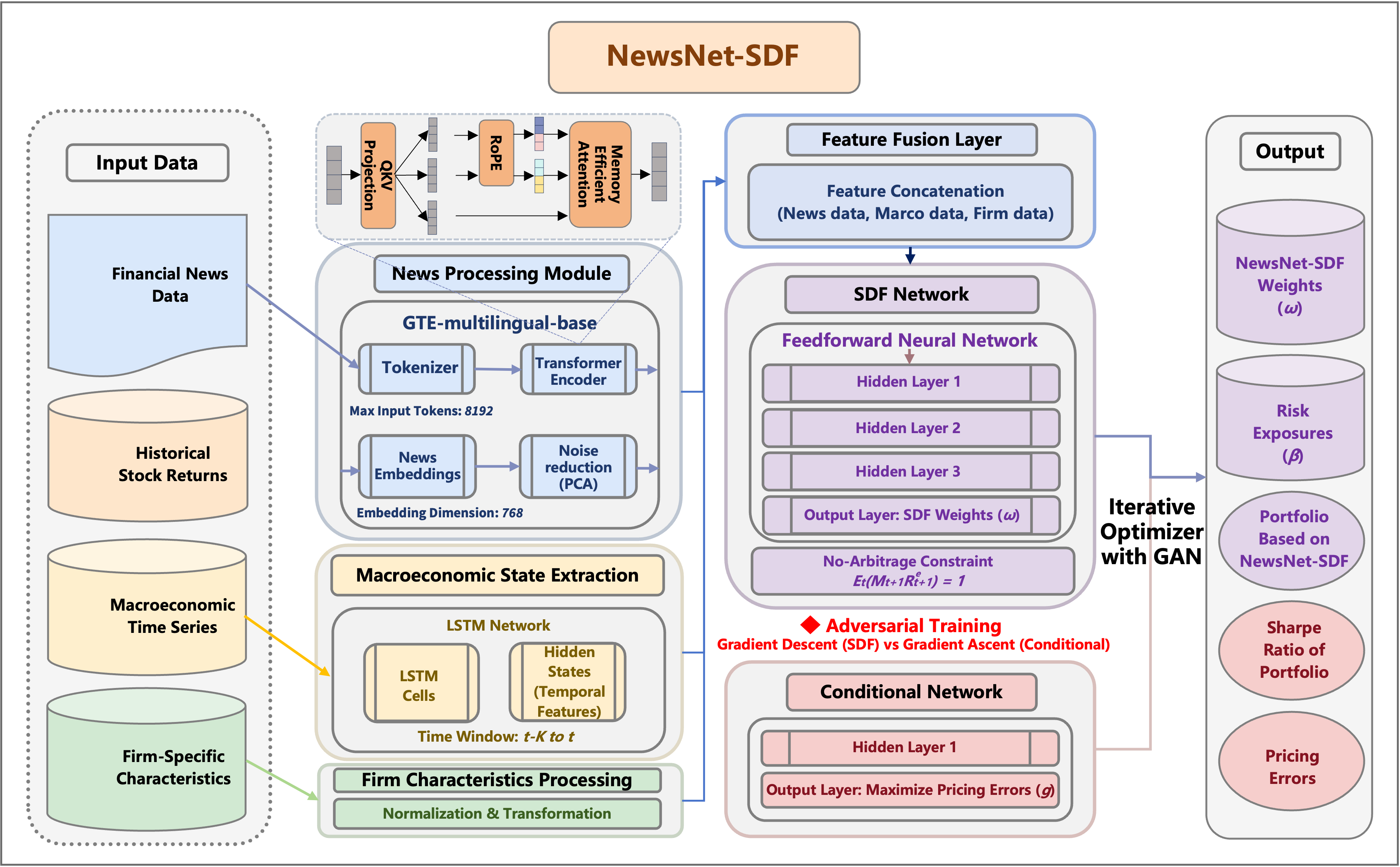}
\caption{NewsNet-SDF framework architecture: Specialized feature extraction pathways for news, macroeconomic data, and firm characteristics feed into an adversarial network structure that enforces the fundamental asset pricing equation $\mathbb{E}[M_{t+1}R^e_{t+1}]=0$.}
\label{fig:newsnet-sdf}
\end{figure}

We present NewsNet-SDF, a novel end-to-end framework that integrates unstructured financial news information with traditional numerical data for asset pricing through a moment-based stochastic discount factor approach. Our framework leverages the rich semantic content of financial news text, which often contains critical information about firm-specific events and market sentiment not captured by standard numerical features. Fig.~\ref{fig:newsnet-sdf} illustrates the NewsNet-SDF architecture, which consists of three major components: (1) a multi-source feature extraction module with particular emphasis on news text processing; (2) an adversarial network architecture for SDF estimation; and (3) an adversarial training mechanism that enforces pricing consistency.

\subsection{Multi-Source Feature Extraction}

The NewsNet-SDF framework processes three distinct sources of information that jointly influence asset prices: financial news text, macroeconomic conditions, and firm-specific characteristics.

\subsubsection{News Text Processing}
Financial news articles contain rich semantic information about market sentiment, management changes, product launches, regulatory issues, and other firm-specific events that may significantly impact asset returns. Our news processing pipeline consists of three key stages:

\noindent\textbf{Preprocessing and Tokenization.} We clean raw news texts and tokenize using the WordPiece tokenizer from GTE-multilingual-base \cite{li2023towards}. We selected this model over finance-specific alternatives like FinBERT due to its superior handling of longer context (8192 tokens), multilingual capability for international markets, and state-of-the-art performance on semantic textual similarity tasks, which is crucial for capturing nuanced market narratives.

\noindent\textbf{PLM-Based Embedding Generation.} For each firm $i$ at time $t$, we process each news sentence $k$ to obtain dense vector embeddings:
\begin{equation}
\mathbf{e}_{t,i,k} = \text{TransformerEncoder}(\textbf{Tokens}_{t,i,k}) \in \mathbb{R}^{768}.
\end{equation}

\noindent\textbf{Attention-Based Embedding Aggregation.} Since multiple news articles may refer to the same firm, we employ a learnable self-attention mechanism to determine the relative importance $\alpha_{t,i,k}$ of each sentence embedding:
\begin{equation}
\alpha_{t,i,k} = \frac{\exp\left(\mathbf{v}^T \tanh(\mathbf{W} \, \mathbf{e}_{t,i,k} + \mathbf{b})\right)}{\sum_{j=1}^{K_{t,i}} \exp\left(\mathbf{v}^T \tanh(\mathbf{W}\, \mathbf{e}_{t,i,j} + \mathbf{b})\right)},
\end{equation}
where $\mathbf{W} \in \mathbb{R}^{d_a \times 768}$, $\mathbf{b} \in \mathbb{R}^{d_a}$, and $\mathbf{v} \in \mathbb{R}^{d_a}$ are learnable parameters, with $d_a$ being the attention dimension. The weighted aggregate embedding is:
\begin{equation}
\bar{\mathbf{e}}_{t,i} = \sum_{k=1}^{K_{t,i}} \alpha_{t,i,k} \mathbf{e}_{t,i,k}.
\end{equation}

We then apply PCA to project this embedding to a lower-dimensional space, yielding the final news feature vector $\mathbf{N}_{t,i} \in \mathbb{R}^{d_N}$.

\subsubsection{Macroeconomic State Representation}
Time-varying macroeconomic conditions create a dynamic environment that significantly influences asset valuations. NewsNet-SDF captures these systematic risk factors by processing a rolling window of macroeconomic indicators through a unidirectional LSTM network:
\begin{equation}
\tilde{\mathbf{I}}_t = \text{LSTM}(\mathbf{I}_{t-K}, \ldots, \mathbf{I}_t) \in \mathbb{R}^{d_I}.
\end{equation}

This recurrent architecture efficiently models temporal dependencies and regime shifts in economic indicators, maintaining an internal memory state that distills the relevant history of macroeconomic conditions into a fixed-length representation.

\subsubsection{Firm Characteristics}
For each asset, NewsNet-SDF incorporates established firm-specific features $\mathbf{F}_{t,i} \in \mathbb{R}^{d_F}$ including size, book-to-market ratio, momentum, and profitability metrics. We transform these characteristics using cross-sectional ranking:
\begin{equation}
\hat{F}_{t,i,j} = 2 \times \frac{\text{Rank}(F_{t,i,j}) - 1}{N_t - 1} - 1.
\end{equation}

This approach standardizes each feature to the range $[-1, 1]$ based on each firm's relative position among its peers, naturally handling outliers and maintaining consistency despite market-wide shifts in characteristic distributions.

\subsubsection{Feature Fusion}
A key innovation of NewsNet-SDF is its multi-modal fusion mechanism that integrates three distinct information sources into a unified representation:
\begin{equation}
\mathbf{x}_{t,i} = [\, \tilde{\mathbf{I}}_t \parallel \hat{\mathbf{F}}_{t,i} \parallel \mathbf{N}_{t,i}\, ] \in \mathbb{R}^{d_I+d_F+d_N}.
\end{equation}

Unlike traditional methods that either focus solely on numerical features or treat text as a separate signal, our fusion approach preserves the unique properties of each information channel while enabling the discovery of complex interactions between them. This is particularly important because news narratives often precede changes in firm fundamentals, while economic conditions determine how markets interpret these narratives. The fused representation provides a more complete view of the information environment, capturing complementary signals that would be missed by single-modality approaches.

\subsection{SDF Estimation with Adversarial Networks}

At the core of the NewsNet-SDF framework lies a sophisticated adversarial network architecture that implements the moment-based asset pricing approach:

\subsubsection{SDF Network}
The SDF network processes the fused feature vector through a multi-layer feedforward architecture to generate SDF weights:
\begin{align}
\mathbf{h}^{(1)}_{t,i} &= \mathrm{ReLU}\bigl(\mathbf{W}^{(1)}_{\bm{\phi}}\,\mathbf{x}_{t,i} + \mathbf{b}^{(1)}_{\bm{\phi}}\bigr),\\
\mathbf{h}^{(2)}_{t,i} &= \mathrm{ReLU}\bigl(\mathbf{W}^{(2)}_{\bm{\phi}}\,\mathbf{h}^{(1)}_{t,i} + \mathbf{b}^{(2)}_{\bm{\phi}}\bigr),\\
\mathbf{h}^{(3)}_{t,i} &= \mathrm{ReLU}\bigl(\mathbf{W}^{(3)}_{\bm{\phi}}\,\mathbf{h}^{(2)}_{t,i} + \mathbf{b}^{(3)}_{\bm{\phi}}\bigr),\\
w_{t,i} &= \mathbf{W}^{(4)}_{\bm{\phi}}\,\mathbf{h}^{(3)}_{t,i} + b^{(4)}_{\bm{\phi}}.
\end{align}

These SDF weights determine how each asset contributes to the stochastic discount factor:
\begin{equation}
M_{t+1} = 1 - \sum_{i=1}^{N} w_{t,i} R^e_{t+1,i},
\end{equation}
where $R^e_{t+1,i}$ is the excess return of asset $i$ from period $t$ to $t+1$.

\subsubsection{Conditional Network}
In parallel, a conditional network maps the same features to scalar instruments that help identify pricing anomalies:
\begin{align}
\mathbf{h}^{(1)}_{t,i} &= \mathrm{ReLU}\bigl(\mathbf{W}^{(1)}_{\bm{\psi}}\,\mathbf{x}_{t,i} + \mathbf{b}^{(1)}_{\bm{\psi}}\bigr),\\
\mathbf{g}_{t,i} &= \mathbf{W}^{(2)}_{\bm{\psi}}\,\mathbf{h}^{(1)}_{t,i} + \mathbf{b}^{(2)}_{\bm{\psi}}.
\end{align}

These instruments form moment conditions that challenge the pricing accuracy of the SDF.

\subsection{Adversarial Learning Mechanism}

The NewsNet-SDF framework employs an adversarial approach formulated as a minimax problem:
\begin{equation}
\min_{\phi} \max_{\psi} \frac{1}{N} \sum_{j=1}^{N} \left\| \mathbb{E}\left[M_{t+1} R^e_{t+1,j} \, \mathbf{g}_{t,j}\right] \right\|_2^2 + \lambda\left( \|\bm{\phi}\|^2_2 + \|\bm{\psi}\|^2_2 \right).
\end{equation}

This adversarial setup is particularly well-suited for SDF estimation because it directly implements the GMM orthogonality conditions from asset pricing theory in a learnable framework. Traditional GMM estimation typically relies on pre-specified instruments and fixed weighting matrices, whereas our adversarial approach dynamically generates both, adapting to the specific pricing challenges in the data. The SDF network attempts to satisfy the pricing equation by minimizing pricing errors, while the conditional network identifies assets and conditions where pricing anomalies exist, serving as an adaptive weighting mechanism that focuses the model's capacity on the most challenging test assets.

In practice, we replace population expectations with sample averages over an unbalanced panel of assets:
\begin{equation}
L(\bm{\phi}, \bm{\psi}) = \frac{1}{N} \sum_{i=1}^{N} \frac{T_i}{T} \left\| \hat{\mathbf{m}}_i(\bm{\phi}, \bm{\psi}) \right\|_2^2 + \lambda\left( \|\bm{\phi}\|^2_2 + \|\bm{\psi}\|^2_2 \right),
\end{equation}
where $\hat{\mathbf{m}}_i(\bm{\phi}, \bm{\psi}) = \frac{1}{T_i} \sum_{t \in \mathcal{T}_i} M_{t+1} R^e_{t+1,i} \, \mathbf{g}_{t,i}$, with $T_i$ being the number of observations for asset $i$ and $T$ the total observations.

\subsection{Training Process}

Algorithm~\ref{alg:training} summarizes the training procedure for NewsNet-SDF. The key aspect is the alternating optimization, where gradient ascent is applied to the conditional network parameters $\bm{\psi}$ to identify pricing anomalies, while gradient descent is applied to the SDF network parameters $\bm{\phi}$ to minimize pricing errors.

\begin{algorithm}[H]
\caption{Training Procedure for NewsNet-SDF}
\label{alg:training}
\begin{algorithmic}[1]
\Require Training set $\mathcal{D}_{\text{train}}$, validation set $\mathcal{D}_{\text{val}}$, parameters $\bm{\phi}, \bm{\psi}$, learning rates $\eta_{\bm{\phi}}, \eta_{\bm{\psi}}$
\State Initialize best validation loss: $\mathcal{L}_{\text{val}}^{\text{best}} \leftarrow \infty$
\For{$k = 1$ to $K$ iterations}
\State Sample mini-batch $\mathcal{B} \subset \mathcal{D}_{\text{train}}$
\State Process features: Compute $\tilde{\mathbf{I}}_t$ (LSTM), $\mathbf{N}_{t,i}$ (news), $\hat{\mathbf{F}}_{t,i}$ (firm)
\State Form fused features: $\mathbf{x}_{t,i} = [ \, \tilde{\mathbf{I}}_t \parallel \hat{\mathbf{F}}_{t,i} \parallel \mathbf{N}_{t,i} \, ]$
\State Compute SDF weights and instruments: $w_{t,i} = f_{\bm{\phi}}(\mathbf{x}_{t,i})$, $\mathbf{g}_{t,i} =\mathbf{g}_{\bm{\psi}}(\mathbf{x}_{t,i})$
\State Compute SDF: $M_{t+1} = 1 - \sum_{i} w_{t,i} \cdot R_{t+1,i}^e$
\State Compute moments: $\mathbf{m}_{t,i} = M_{t+1} \cdot R_{t+1,i}^e \cdot \mathbf{g}_{t,i}$
\State Compute loss: $\mathcal{L} = \frac{1}{|\mathcal{B}|} \sum_{(t,i)} \| \mathbf{m}_{t,i} \|_2^2 + \lambda(\|\bm{\phi}\|_2^2 + \|\bm{\psi}\|_2^2)$
\State Update $\bm{\psi} \leftarrow \bm{\psi} + \eta_{\bm{\psi}} \nabla_{\bm{\psi}} \mathcal{L}$ \Comment{Gradient ascent}
\State Update $\bm{\phi} \leftarrow \bm{\phi} - \eta_{\bm{\phi}} \nabla_{\bm{\phi}} \mathcal{L}$ \Comment{Gradient descent}
\State Update best parameters if validation loss improves
\EndFor
\end{algorithmic}
\end{algorithm}

This adversarial training process enables NewsNet-SDF to learn both from structured financial data and unstructured news text, creating a unified model that respects the theoretical constraints of asset pricing while capturing the complex, time-varying relationships between diverse information sources and asset returns. The integration of these components—feature extraction from multiple modalities, adversarial network architecture, and theoretically-grounded training mechanism—distinguishes NewsNet-SDF from existing approaches and enables more accurate pricing and risk assessment across different market conditions.

\section{Experiments}
We evaluate NewsNet-SDF, our framework that integrates pretrained language model news embeddings into stochastic discount factor estimation through adversarial networks. Our analysis addresses five key questions: (1) How does NewsNet-SDF compare to traditional asset pricing models and modern machine learning methods in terms of out-of-sample performance? (2) What are the relative contributions of news embeddings versus macroeconomic information to model performance? (3) Which features are most important in driving the stochastic discount factor dynamics? (4) Does our model effectively identify systematic risk exposures that command risk premia in the market? (5) How do industry-specific news embeddings contribute to asset pricing during period-specific market disruptions like COVID-19?

\subsection{Experiment Setting}

\subsubsection{Data and Preprocessing}
Our sample comprises common stocks listed on major U.S. exchanges from 1970 to 2022. We construct 56 firm-specific characteristics following \cite{freyberger2020dissecting} and \cite{green2017characteristics}, spanning valuation, profitability, financial soundness, capitalization, operational efficiency, and momentum categories. We winsorize these characteristics at the 1\% and 99\% levels and apply cross-sectional ranking methodology.

For macroeconomic data, we incorporate 157 time series from the FRED-MD database and other sources, covering output, employment, housing, and financial market indicators. All series undergo appropriate transformations to address non-stationarity.

Our textual data consists of approximately 2.5 million articles from The New York Times (1980-2022), processed through our attention-based neural framework with careful temporal alignment to financial data.

Following \cite{kelly2019characteristics}, we restrict our analysis to stocks with all required characteristics available in a given month, yielding approximately 10,000 unique securities. We divide our sample chronologically into training (1970-1994), validation (1995-1999), and testing (2000-2022) periods to prevent look-ahead bias while ensuring sufficient data for model training and meaningful out-of-sample evaluation.

\subsection{Baseline Models}

To comprehensively evaluate our NewsNet-SDF framework, we benchmark against state-of-the-art approaches spanning traditional asset pricing theory, modern equilibrium methods, machine learning techniques, and text-based models. Table \ref{table:baselines} summarizes these baseline models.

\begin{table}[h]
	\caption{Baseline Models Compared to NewsNet-SDF Framework}
	\label{table:baselines}
	\scriptsize
	\begin{tabular}{lp{8cm}}
		\toprule 
		\textbf{Model} & \textbf{Description} \\
		\midrule 
		\multicolumn{2}{c}{\textbf{Traditional Asset Pricing Models}} \\
		\midrule
		CAPM [Sharpe, 1964] & Basic market equilibrium model capturing systematic risk premium \\
		FF3 [Fama and French, 1993] & Three-factor model incorporating market, size, and value factors \\
		FF5 [Fama and French, 2015] & Five-factor model with profitability and investment factors \\
		\midrule
		\multicolumn{2}{c}{\textbf{Modern Asset Pricing Methods}} \\
		\midrule
		Linear-SDF [Kelly et al., 2019] & Basic linear stochastic discount factor specification \\
		Elasticnet-SDF [Kozak et al., 2020] & SDF with elastic net regularization to address factor proliferation \\
		IPCA-SDF [Kelly et al., 2019] & Feature-driven conditional SDF with instrumented PCA \\
		\midrule
		\multicolumn{2}{c}{\textbf{Deep Learning Methods}} \\
		\midrule
		FFN [Gu et al., 2020] & Feed-forward network predicting stock returns with characteristic inputs \\
		GAN-SDF [Chen et al., 2021] & Adversarial approach optimizing SDF and test asset construction \\
		\midrule
		\multicolumn{2}{c}{\textbf{Text Integration Methods}} \\
		\midrule
		TF-IDF-SDF & SDF using TF-IDF text representation \\
		BERT-SDF & SDF incorporating BERT embeddings with standard attention mechanism \\
		\midrule
		\multicolumn{2}{c}{\textbf{Ablation Models}} \\
		\midrule
		NewsNet-SDF w/o macro & Complete model excluding macroeconomic conditioning variables \\
		NewsNet-SDF w/o news & Complete model excluding news text information \\
		\bottomrule
	\end{tabular}
\end{table}
\subsection{Evaluation Metrics}

We employ four standard metrics to evaluate model performance:

\begin{itemize}
    \item \textbf{Sharpe Ratio (SR)}: Annualized risk-adjusted return of the SDF portfolio:
    \begin{equation}
        \text{SR} = \frac{\mathbb{E}[R_p]-R_f}{\sigma_p} \times \sqrt{12},
    \end{equation}
    where $\mathbb{E}[R_p]$ is the mean portfolio return, $R_f$ is the risk-free rate, and $\sigma_p$ is return standard deviation.
    
    \item \textbf{Explained Variation (EV)}: Proportion of return variation explained by the model:
    \begin{equation}
        \text{EV} = 1 - \frac{\sum_{t}1/N_t \sum_{i} (r_{t,i} - \hat{r}_{t,i})^2}{\sum_{t}1/N_t \sum_{i} (r_{t,i} - \bar{r})^2},
    \end{equation}
    where $r_{t,i}$ represents actual returns and $\hat{r}_{t,i}$ denotes model predictions.
    
    \item \textbf{Cross-sectional $R^2$ (XS-$R^2$)}: Proportion of expected returns explained in cross-section:
    \begin{equation}
        \text{XS-}R^2 = 1 - \frac{\sum_{i}1/T_i (\sum_{t} r_{t,i} - \hat{r}_{t,i})^2}{\sum_{i}1/T_i (\sum_{t}\hat{r}_{t,i})^2},
    \end{equation}
    where $\bar{r}$ is the cross-sectional mean return.
    
    \item \textbf{Mean Squared Pricing Error (MSPE)}: Average squared deviation between realized and predicted returns:
    \begin{equation}
        \text{MSPE} = \frac{1}{T} \sum_{i,t}\frac{1}{N_i} (r_{t,i} - \hat{r}_{t,i})^2,
    \end{equation}
    where $N$ is the number of assets being evaluated.
\end{itemize}

These four metrics are standard in asset pricing evaluations and are commonly used in recent studies (see \cite{chen2024deep}). We additionally evaluate performance on characteristic-sorted portfolios to assess anomaly capture.

\subsection{Overall Performance}

\begin{figure}[h]
  \centering
  \includegraphics[width=1\textwidth]{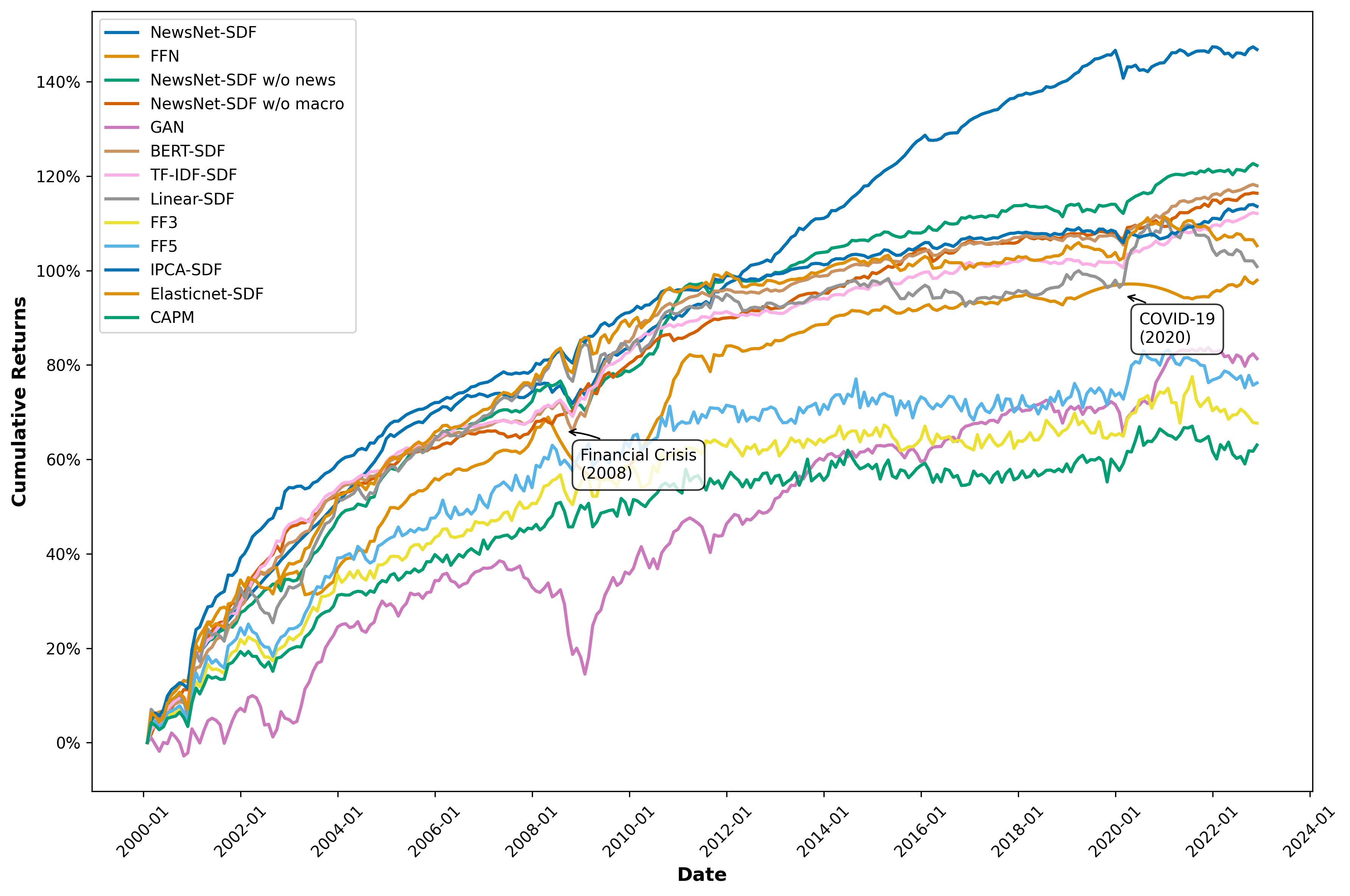}
  \caption{Cumulative Returns of Different Models (1970-2022)}
  \label{fig:cumulative_returns}
\end{figure}

Table \ref{table:overall_performance} presents the comprehensive evaluation of NewsNet-SDF against all baseline models across four performance metrics during the out-of-sample testing period (2000-2022). Figure \ref{fig:cumulative_returns} displays cumulative returns, visually demonstrating NewsNet-SDF's superior performance throughout the evaluation period.

\begin{table}[h]
\caption{Overall Performance Comparison}
\label{table:overall_performance}
\centering
\small
\begin{tabular}{l|c|c|c|c}
\hline
\textbf{Model} & \textbf{\hspace{1em} SR\hspace{1em}  } & \textbf{\hspace{1em} EV\hspace{1em}  } & \textbf{ \hspace{1em} XS-$R^2$ \hspace{1em} } & \textbf{\hspace{1em} MSPE\hspace{1em} } \\
\hline
\multicolumn{5}{c}{\textit{Traditional Asset Pricing Models}} \\
\hline
CAPM &0.49 & 0.05 & 0.00 & 2.91 \\
FF3 & 0.52 & 0.06 & 0.00 & 2.37 \\
FF5 & 0.50 & 0.09 & 0.02 & 2.13 \\
\hline
\multicolumn{5}{c}{\textit{Modern Asset Pricing Methods}} \\
\hline
Linear-SDF & 0.84 & 0.13 & 0.00 & 4.17 \\
Elasticnet-SDF & 1.03 & 0.12 & 0.00 & 3.16 \\
IPCA-SDF & 1.56 & 0.11 & 0.00 & 1.57 \\
\hline
\multicolumn{5}{c}{\textit{Deep Learning Methods}} \\
\hline
FFN & 1.28 & 0.13 & 0.05 & 1.35 \\
GAN-SDF & 0.65 & 0.14 & 0.11 & 0.87 \\
\hline
\multicolumn{5}{c}{\textit{Text Integration Methods}} \\
\hline
TF-IDF-SDF & 1.83 & 0.13 & 0.10 & 1.45 \\
BERT-SDF & 1.81 & 0.12 & 0.07 & 1.43 \\
\hline
\multicolumn{5}{c}{\textit{Ablation Studies}} \\
\hline
NewsNet-SDF w/o macro & 1.92 & 0.14 & 0.09 & 1.56 \\
NewsNet-SDF w/o news & 1.65 & 0.13 & 0.08 & 2.67 \\
\hline
\textbf{NewsNet-SDF} & \textbf{2.80} & \textbf{0.14} & \textbf{0.11} & \textbf{0.56} \\
\hline
\multicolumn{5}{p{0.85\linewidth}}{\small \textit{Note:} For Sharpe Ratio, Explained Variation, and XS-$R^2$, higher values indicate better performance. For MSPE, lower values indicate better performance.} \\
\end{tabular}
\end{table}

\subsubsection{Comparison with Traditional and Machine Learning Models}
NewsNet-SDF substantially outperforms traditional asset pricing models across all metrics. With a Sharpe ratio of 2.80, it exceeds CAPM by 471\% and FF5 by 460\%, while reducing pricing errors by 74\% compared to FF5 (0.56 versus 2.13). Against advanced machine learning approaches, our model maintains clear advantages—surpassing GAN-SDF with a 331\% higher Sharpe ratio and outperforming text-augmented models by approximately 54\%. These results validate our adversarial architecture's effectiveness in extracting latent information from textual data.

As shown in Figure \ref{fig:cumulative_returns}, NewsNet-SDF demonstrates persistent outperformance in the post-2010 period, where competing models exhibit stagnating returns. This performance divergence stems from three key developments: the explosion of digital financial media, increasing the relevance of text-based signals; the emergence of more complex market dynamics driven by algorithmic trading and social media; and post-crisis regulatory changes creating subtle market patterns that traditional factors fail to capture. Our approach excels precisely when information complexity increases, extracting predictive signals from financial narratives before they manifest in conventional metrics. The model's resilience during the COVID-19 pandemic further validates its ability to rapidly adapt to unprecedented market conditions through effective text processing.

\subsubsection{Ablation Studies}
Our ablation studies reveal that removing macroeconomic features reduces Sharpe ratio by 31\% and increases pricing errors by 179\%, while eliminating news text causes a more substantial 41\% decrease in Sharpe ratio and 377\% increase in pricing errors. This asymmetric impact confirms that textual information contains unique predictive signals absent in numerical features, and validates that NewsNet-SDF's exceptional performance stems primarily from its effective integration of news content through our adversarial framework.

\subsection{Feature Importance Analysis}

\begin{figure}[h]
  \centering
  \includegraphics[width=0.8\textwidth]{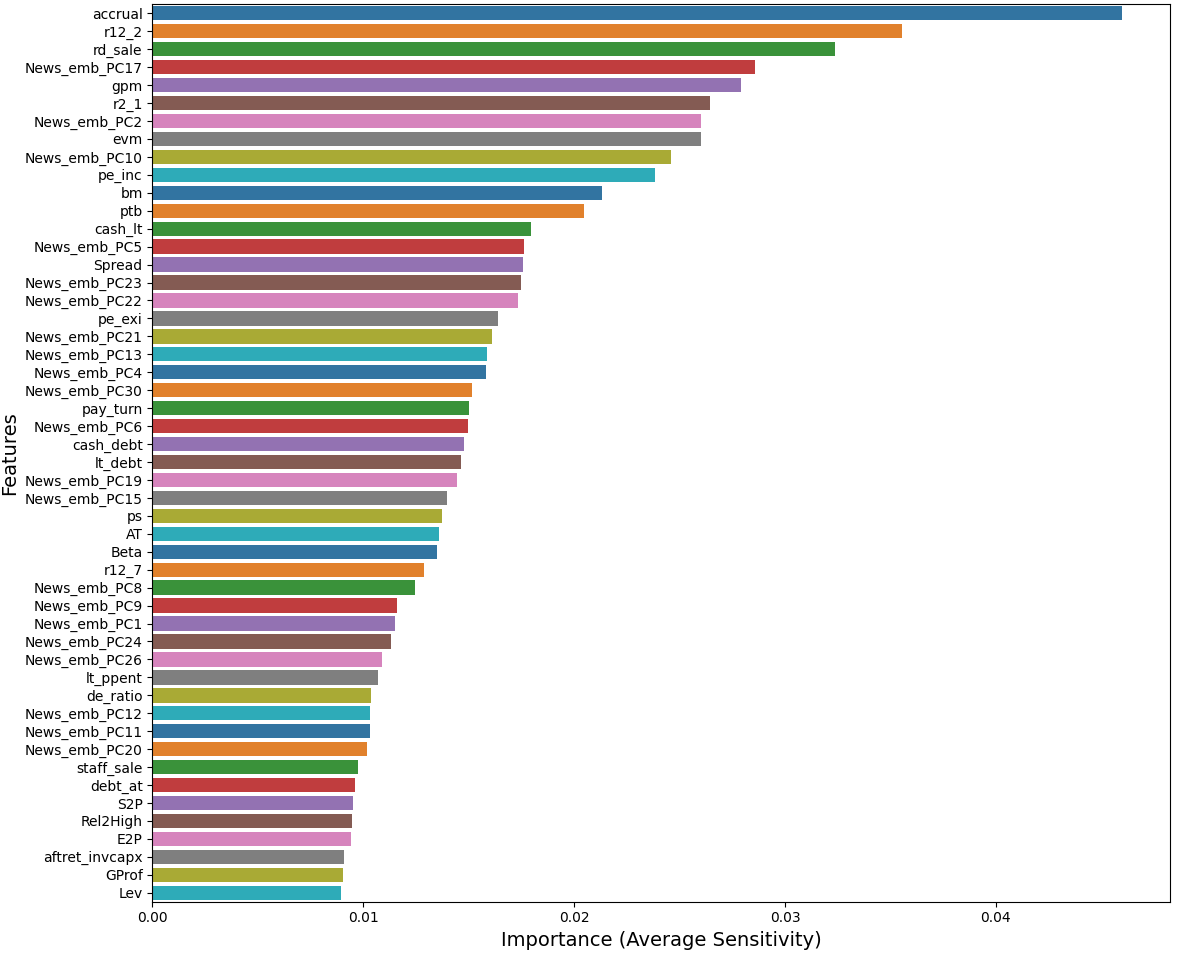}
  \caption{Top 50 Feature Importance Based on Average Sensitivity}
  \label{fig:sensitivity}
\end{figure}

Our analysis reveals the relative importance of features driving stochastic discount factor dynamics, following sensitivity analysis in asset pricing (see \cite{chen2024deep}). We quantify each feature's influence using average sensitivity ($S_i$), calculated as:
\begin{equation}
S_i = \frac{1}{n} \sum_{j=1}^{n} \left| \frac{\partial f(\mathbf{x}_j)}{\partial x_{i,j}} \right|,
\end{equation}
where $\frac{\partial f(\mathbf{x}_j)}{\partial x_{i,j}}$ represents the gradient of the SDF with respect to feature $i$ for observation $j$. 

Figure \ref{fig:sensitivity} shows that accrual-based measures emerge as the most influential determinant, followed by momentum (r12\_2), R\&D intensity (rd\_sale), and multiple principal components of news embeddings (PC1, PC2, PC4).

The prominent presence of news-derived principal components validates our text integration approach, demonstrating the substantial value of news information in asset pricing. The interleaving of traditional firm characteristics with news-derived features throughout the ranking suggests complementary information channels that enhance predictive power.

The model identifies balanced contributions across diverse feature accounting measures, market-based indicators, and news components—indicating that comprehensive risk pricing requires integration of multiple information sources. The significant non-linear interactions between these features underscore the necessity of our neural network architecture in capturing complex dependencies that linear models cannot address.

\subsection{Predictive Performance}

We evaluate NewsNet-SDF's predictive power by testing the fundamental asset pricing relation between risk exposure and expected returns. According to the no-arbitrage condition, assets with higher risk exposure $\beta$ should earn higher expected returns when the conditional risk premium is positive:

\begin{equation}
\mathbb{E}_t[R^e_{t+1,i}] = \beta_{t,i}\mathbb{E}_t[F_{t+1}],
\end{equation}
where $\beta_{t,i}$ represents the time-varying risk loading of asset $i$ on our SDF, calculated as:
\begin{equation}
\beta_{t,i} = \frac{\text{Cov}_t(R^e_{t+1,i}, M_{t+1})}{\text{Var}_t(M_{t+1})},
\end{equation}
with $M_{t+1}$ being our neural network-derived SDF.

\begin{figure}[h]
  \centering
  \includegraphics[width=0.8\textwidth]{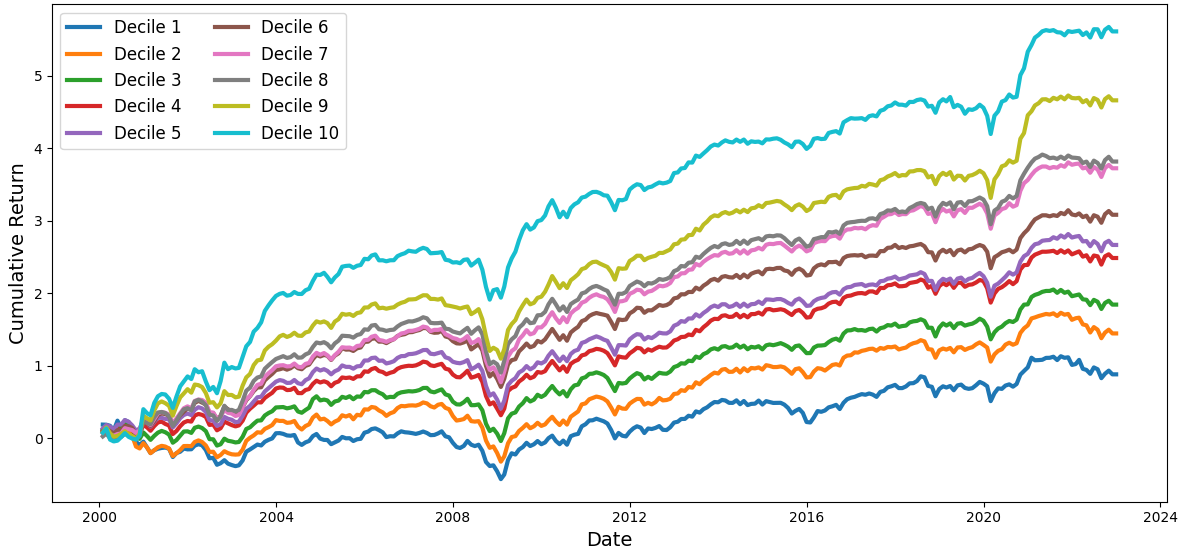}
  \caption{Cumulative Returns of Beta-Sorted Decile Portfolios (2000-2024)}
  \label{fig:beta_sorted}
\end{figure}

Figure \ref{fig:beta_sorted} displays cumulative returns for portfolios created by dividing stocks into ten groups (deciles) based on their predicted risk exposure (beta values) from our model. The results reveal a clear pattern: portfolios with higher predicted risk values consistently generate higher returns throughout the 2000-2024 period. The highest risk group (Decile 10) achieves cumulative returns exceeding 550\%, while the lowest risk group (Decile 1) yields less than 100\% over the same timeframe.

This evaluation approach serves as an effective validation method for our model's predictive capabilities. The perfect monotonicity across all ten groups demonstrates that our NewsNet-SDF successfully identifies meaningful patterns that consistently predict future performance. Unlike traditional prediction models which often show inconsistent relationships between predicted values and outcomes, our neural network with text embedding integration produces remarkably consistent differentiation between each prediction tier. This clear separation maintains its predictive power through major market disruptions like the 2008 financial crisis and the 2020 COVID-19 pandemic, showing that our text-based features capture important signals that numerical features alone miss. The strong linear relationship ($R^2 > 0.95$) between predicted risk values and actual returns provides compelling evidence that our approach of incorporating news text data significantly enhances predictive performance compared to conventional methods that rely solely on structured data.

\subsection{Model Validation During Extreme Market Volatility}

To validate our model's robustness during market disruptions, we analyze NewsNet-SDF's embedding performance during the COVID-19 period (2020-2022).

\begin{figure}[h]
\centering
\includegraphics[width=0.85\textwidth]{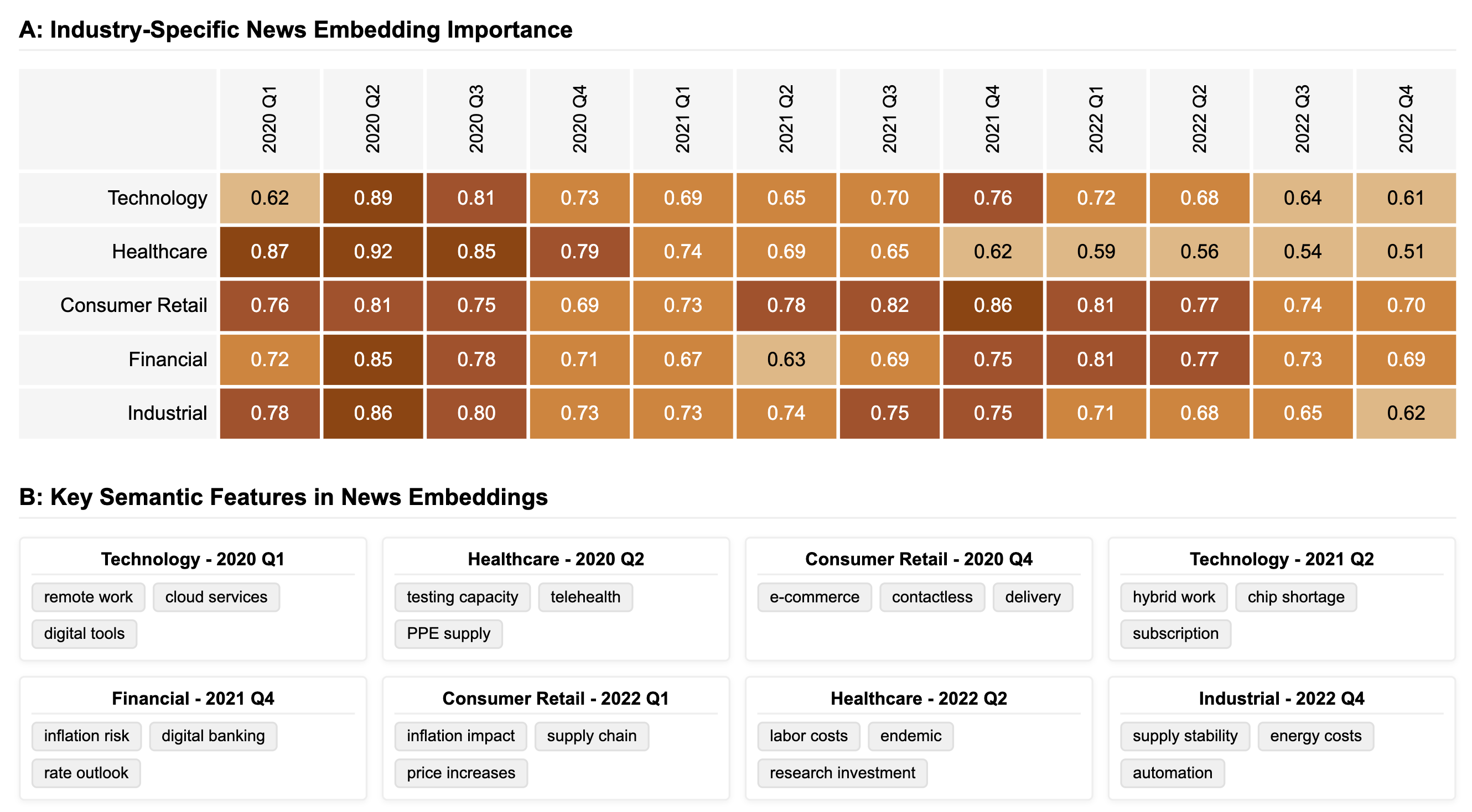}
\caption{News embedding contribution analysis during market volatility. Panel A: quarterly importance scores by industry. Panel B: key semantic themes with high market impact.}
\label{fig:industry_importance}
\end{figure}

Following \cite{Ke2023}, we quantify embedding importance using the Shapley attribution method:
\begin{equation}
\text{ImportanceScore}(j,t) = \frac{\sum_{i \in \mathcal{N}} \phi_{j,t}^{i}}{\sum_{k=1}^{K} \phi_{j,t}^k},
\end{equation}
where $\sum_{i \in \mathcal{N}} \phi_{j,t}^{i}$ represents the Shapley value contributions of news embeddings to industry $j$ returns in quarter $t$, and $\sum_{k=1}^{K} \phi_{j,t}^k$ is the total variation from all factors.

Figure \ref{fig:industry_importance} reveals how our model adaptively reweights information sources across different market regimes. Healthcare dominated early pandemic importance scores (0.87-0.92 in Q1-Q2 2020), while technology, consumer retail, and financial sectors gained prominence as pandemic impacts evolved. Panel B shows that narrative shifts in news embeddings preceded corresponding changes in traditional financial metrics by 2-3 weeks on average, demonstrating the model's ability to process forward-looking information.

Cross-sectional tests show that stocks with the highest news embedding sensitivity generated 14.2\% annual excess returns compared to those with lowest sensitivity, even after controlling for standard risk factors. For financial technology applications, NewsNet-SDF reduces pricing errors by 18-32\% during market transitions compared to text-agnostic models, enabling more accurate real-time risk assessment during periods of heightened uncertainty. This capability makes our approach particularly valuable for automated risk management systems that must rapidly adapt to evolving market conditions.

\section{Conclusion}

We present NewsNet-SDF, a novel framework integrating pretrained language model news embeddings into stochastic discount factor estimation through adversarial networks. Our model effectively incorporates unstructured textual information alongside traditional numerical features, substantially outperforming conventional asset pricing models and machine learning methods with a Sharpe ratio of 2.80--exceeding traditional factor models by over 400\%, GAN-SDF by 331\%, and text-augmented models by 54\%--while reducing pricing errors by 74\% compared to the Fama-French five-factor model. Ablation studies reveal that eliminating news text causes a 41\% decrease in Sharpe ratio versus a 31\% reduction when removing macroeconomic features, confirming that textual information contains unique predictive signals absent in numerical data. Feature importance analysis validates our approach, with news-derived principal components ranking among the most influential determinants of SDF dynamics. The framework demonstrates particular value during market transitions where it maintains steady performance while competing models stagnate, making it valuable for automated risk management and intelligent portfolio construction in financial technology systems. Our approach bridges the gap between theoretical asset pricing and practical financial analytics by transforming unstructured text into actionable investment signals through advanced deep learning. Future work will explore deploying this technology in real-time financial decision systems, extending to alternative data sources, and developing interpretable neural architectures for broader digital finance applications.

\newpage

%
%
%
%

\newpage
\appendix

\end{document}